\documentclass[11pt,a4paper]{article}
\usepackage[hyperref]{naaclhlt2019}
\usepackage{times}
\usepackage{latexsym}
\usepackage{amssymb}
\usepackage{amsmath}
\usepackage{graphicx}
\usepackage{url}
\usepackage{bm}

\aclfinalcopy 

\title{The Verbal and Non Verbal Signals of Depression - Combining Acoustics, Text and Visuals for Estimating Depression Level}

\author{Syed Arbaaz Qureshi\\
  Indian Institute of Technology Patna \\
  Patna, India \\
  \texttt{arbaaz.qureshi29@gmail.com} \\\And
  Mohammed Hasanuzzaman \\
  ADAPT Centre \\
  Dublin, Ireland \\
  \texttt{hasanuzzaman.im@gmail.com} \\\AND
  Sriparna Saha \\
  Indian Institute of Technology Patna \\
  Patna, India \\
  \texttt{sriparna@iitp.ac.in} \\\And
  Ga{\"e}l Dias \\
  University of Caen Normandy\\
  Caen, France \\
  \texttt{gael.dias@unicaen.fr}\\
  }

\begin{document}
\maketitle
\begin{abstract}
Depression is a serious medical condition that is suffered by a large number of people around the world. It significantly affects the way one feels, causing a persistent lowering of mood. In this paper, we propose a novel attention-based deep neural network which facilitates the fusion of various modalities. We use this network to regress depression level. Acoustic, text and visual modalities have been used to train our proposed network. Various experiments have been carried out on the benchmark dataset, namely, Distress Analysis Interview Corpus - a Wizard of Oz (DAIC-WOZ). From the results, we empirically justify that the fusion of all three modalities helps in giving the most accurate estimation of depression level. Our proposed approach outperforms the state-of-the-art by 7.17\% on root mean squared error (RMSE) and 8.08\% on mean absolute error (MAE).
\end{abstract}


\section{Introduction}

Depression is a common and serious medical illness that negatively affects how one feels. It is characterized by persistent sadness, loss of interest and an inability to carry out activities that one normally enjoys. It is the leading cause of ill health and disability worldwide. More than 300 million people are now living with depression, an increase of more than 18\% between 2005 and 2015.\footnote{A statistic reported by the World Health Organization available at \url{https://bit.ly/2rsqQoP}.}

Depression lasts between 4 and 8 months on average, and a few of the symptoms and side effects of depression are insomnia, weight loss, fatigue, feelings of worthlessness, drug or alcohol abuse, and impaired ability to think, concentrate and make decisions. In extreme cases, it may also be characterized by thoughts of death, suicide and attempt of suicide. Tragically, the annual number of death cases due to depression is on the rise.\footnote{A study by Hannah Ritchie and Max Roser in 2018 available at \url{https://bit.ly/2mnyVZ6}.}

The causes of depression are not completely known and they may not be down to a single source. Major depressive disorder is likely to be due to complex combinations of factors like genetics, psychology, and social surroundings of the sufferer. People who have experienced life events like divorce or death of a family member or friend, people who have personality issues such as the inability to deal with failure and rejection, people with previous records of major depression, and people with childhood trauma are at a higher risk of depression \cite{beck2009depression}.

Depression detection is a challenging problem as many of its symptoms are covert. Since depressed people socialize less, its detection becomes difficult. Today, for the correct diagnosis of depression, a patient is evaluated on standard questionnaires. In the literature, different tools for screening depression have been proposed, such as the Personal Health Questionnaire Depression Scale (PHQ), the Hamilton Depression Rating Scale (HDRS), the Beck Depression Inventory (BDI), the Center for Epidemiologic Studies Depression Scale (CES-D), the Hospital Anxiety and Depression Scale (HADS), and the Montgomery and Asberg Depression Rating Scale (MADRS).\footnote{Recommandation of the French Haute Autorit\'e de la Sant\'e available at \url{https://bit.ly/2EaOs92}.}  In particular, the eight-item PHQ-8 \cite{Kroenke2009} is established as a valid diagnostic and severity measure for depressive disorders in many clinical studies \cite{kroenke2012enhancing}.


The steadily increasing global burden of depression and mental illness acts as an impetus for the development of more advanced, personalized and automatic technologies that aid in its detection. Affective computing is one field of research which focuses on gathering data from faces, voices and body language to measure human emotion. An important business goal of affective computing is to build human-computer interfaces that can detect and appropriately respond to an end user's state of mind. As a consequence, techniques from affective computing have been applied for the automatic detection of depression \cite{SchererLGRM16,morales2018linguistically}.

In this paper, we introduce an attention-based neural network for the fusion of the acoustic, text and visual modalities. In particular, we encode seven modalities (two acoustic, one text and four visual). Different combinations of the acoustic, text and visual modality encodings are fed to the network to obtain fused vectors. These fused vectors are then passed to a deep regression network to predict the severity of depression based on a PHQ-8 scale. From our experiments, we show that:
\begin{itemize}
    \item the fusion of all modalities (acoustic, text, visual) helps in better estimation of depression level, compared to any other combination,
    \item our approach outperforms the previous state-of-the-art by 7.17\% on root mean squared error (RMSE) and 8.08\% on mean absolute error (MAE) and,
    \item the verbal input plays a predominant role in the regression process, confirming therapists’ experience.
\end{itemize}

The remainder of this paper is organized as follows. In section 2, we present the state-of-the-art approaches for the estimation of depression level. We then describe our methodology in section 3. This is followed by a brief overview of the multimodal DAIC-WOZ dataset used to benchmark our method, in section 4. Experiments, experimental settings and results are described in section 5. Section 6 concludes the paper.


\section{Related work}

Over the last few years, a great deal of research studies in Computer Science have been proposed to deal with mental health disorders \cite{Andersson2014,Dewan:2015}. Within this context, the automatic detection of depression has received major focus. Some initial initiatives have targeted the understanding of relevant descriptors that could be used in machine learning frameworks. \cite{scherer2013automatic} investigate the capabilities of automatic non verbal behavior descriptors to identify indicators of psychological disorders such as depression. In particular, they propose four descriptors that can be automatically estimated from visual signals: downward angling of the head, eye gaze, duration and intensity of smiles, and self-touches. \cite{chatterjee2014context} study the role of multiple context-based heart-rate variability descriptors for evaluating a
person’s psychological health. \cite{cummins2015review} focus on how common paralinguistic speech characteristics (prosodic features, source features, formant features, spectral features) are affected by depression and suicidality and the application of this information in classification and prediction systems. \cite{morales2016speech} argue that researchers should look beyond the acoustic properties of speech by building features that capture syntactic structure and semantic content. Within this context, \cite{Wolohan2018} show that overall classification performance suggests that lexical models are reasonably robust and well suited for a role in a diagnostic or the monitoring capacity of depression. Some other interesting work directions using text features include the study of social media \cite{de2013predicting,Hovy2017}, eventually using specific corpora tuned for such tasks \cite{losada2016test}.

Another promising research trend aims at leveraging all modalities into one learning model and is commonly called multimodal depression detection \cite{morales2018multimodal}. Within this context, a great deal of successful research studies have been proposed. \cite{he2015multimodal} evaluate feature fusion and model fusion strategies via local linear regression to improve accuracy in the BDI score using visual and acoustic cues. \cite{Dibeklioglu:2015} compare facial movement dynamics, head movement dynamics, and vocal prosody individually and in combination, and show that multimodal measures afford most powerful detection. \cite{Yang:2016} achieve satisfying results over the benchmark dataset, DAIC-WOZ (Distress Analysis Interview Corpus - a Wizard of Oz), to estimate the PHQ-8 score by fusing audio, visual and text features with decision-tree classification. More recently, \cite{morales2018linguistically,morales2018multimodal} propose an extensive study of fusion techniques (early, late and hybrid) for depression detection combining audio, visual and text (especially syntactic) features, through SVM. In particular, they show that the syntax-informed fusion approach is able to leverage syntactic information to target more informative aspects of the speech signal, but the overall results tend to suggest that there is no statistical evidence of this finding.

In this paper, we propose an early fusion strategy using neural networks, to combine acoustic, visual and text modalities. For that purpose, different combinations of modalities are fed to an attention-based neural network to obtain fused vectors. These fused vectors are then passed to a deep regression network to predict the severity of depression based on a PHQ-8 scale over the benchmark dataset, DAIC-WOZ. The main motivation of our work is to automatically learn the significance of a given modality as each one may not have the same discriminative characteristics. For that purpose, attention-based neural networks are particularly suitable models, which is confirmed by the overall results obtained in this paper. To the best of our knowledge, this is the first attempt in that direction.


\section{Methodology}
Our proposed architecture (CombAtt) consists of three main components: (1) modality encoders, which take unimodal features as input, and output modality encodings, (2) the fusion subnetwork that fuses the individual modalities, and (3) the regression subnetwork that outputs the estimated PHQ-8 score, conditioned on the output of the fusion subnetwork. 

Let \textit{TSD} = \{\textit{TSD\textsubscript{1}}, \textit{TSD\textsubscript{2}}, ..., \textit{TSD\textsubscript{m}}\} be a set of \textit{m} modalities of time series data. Each element in \textit{TSD} is a two dimensional matrix. The rows of this matrix comprise time stamped feature vectors. The CombAtt network takes the elements of \textit{TSD} as its input. The following subsections describe the individual components of the CombAtt network in detail.


\subsection{Modality encoders}
There are \textit{m} encoders in the CombAtt network, to encode the \textit{m} modalities of time series data. Each of them encodes a modality into an encoding vector. Let network \textbf{ME\textsubscript{k}} $\in$ \{\textbf{ME\textsubscript{1}}, \textbf{ME\textsubscript{2}}, ..., \textbf{ME\textsubscript{m}}\}, and \textit{W\textsubscript{k}} $\in$ \{\textit{W\textsubscript{1}}, \textit{W\textsubscript{2}}, ..., \textit{W\textsubscript{m}}\} be the set of parameters of the network \textbf{ME\textsubscript{k}}. Let \textit{TSD\textsubscript{k}} $\in$ \textit{TSD} be the input to \textbf{ME\textsubscript{k}}. The encoding \textit{E\textsubscript{k}} $\in$ \{\textit{E\textsubscript{1}}, \textit{E\textsubscript{2}}, ..., \textit{E\textsubscript{m}}\} (= \textit{E}) of the input \textit{TSD\textsubscript{k}}, obtained from \textbf{ME\textsubscript{k}} is given by equation \ref{eq:1}. 
\begin{equation}
    \textit{E\textsubscript{k}} = \textbf{ME\textsubscript{k}} (\textit{TSD\textsubscript{k}} : \textit{W\textsubscript{k}})
    \label{eq:1}
\end{equation}
To encode time series data, we use a LSTM network, with a forget gate \cite{gers1999learning} as the recurrent unit, because of its robustness in capturing long sequences. The formulation of our LSTM network is given in the appendix, in section 2. The output from the LSTM layer acts as the encoding vector for a given modality. These modality encoding vectors are then fed to the fusion network.


\subsection{Fusion subnetwork}
The fusion subnetwork is composed of the tensor fusion layer, which disentangles unimodal and bimodal dynamics, and the attention fusion subnetwork, in which the attention mechanism allows to automatically weight modalities. Both components are described in the following sections.
\subsubsection{\textbf{Tensor fusion layer}}
\begin{figure}
\includegraphics[scale = 0.6]{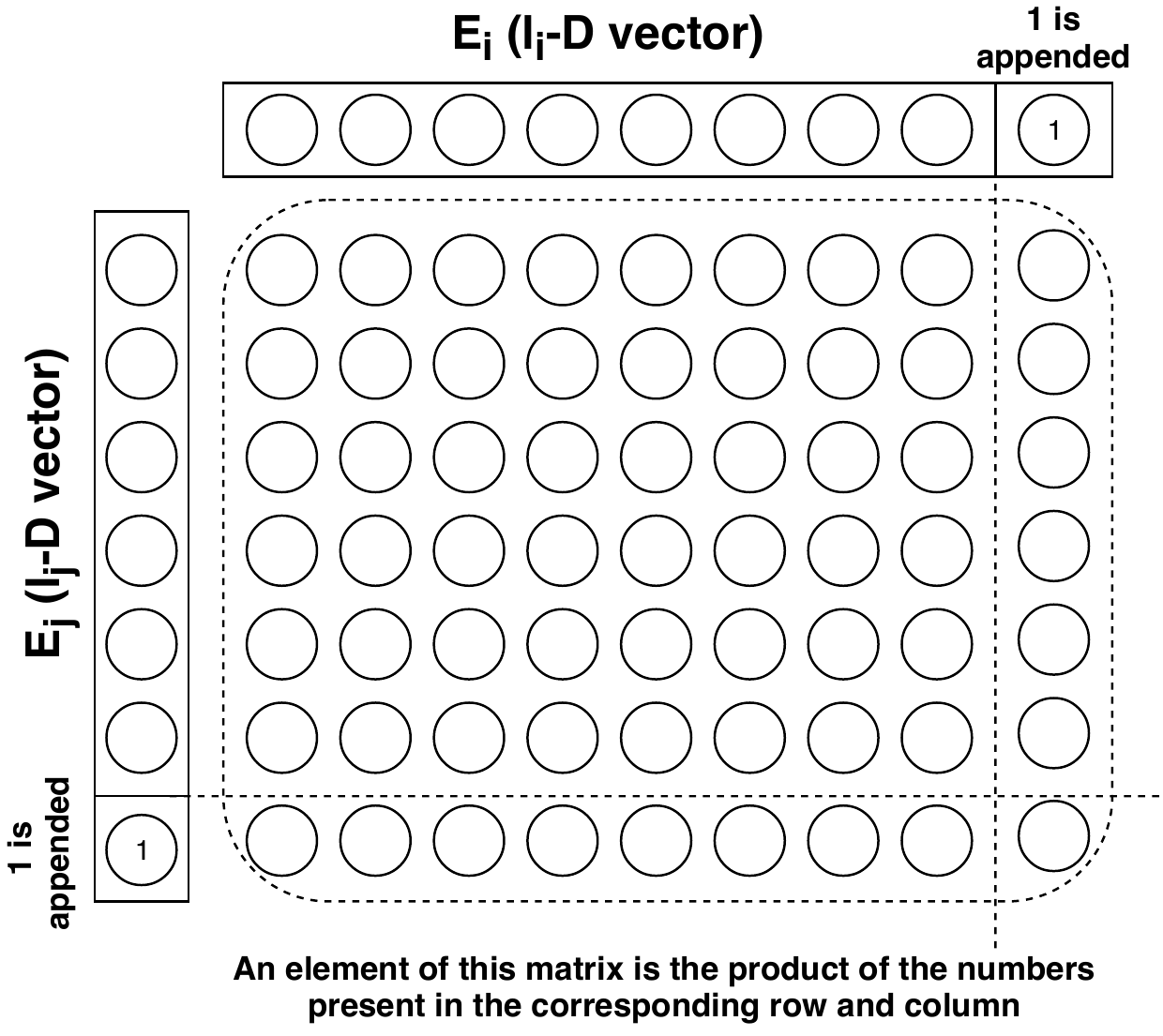}
\caption{Tensor fusion layer.}
\label{fig:1}
\end{figure}

The tensor fusion layer (\textbf{TFL}) disentangles unimodal and bimodal dynamics by modeling each of them explicitly. So, if \textit{E\textsubscript{i}} (\textit{l\textsubscript{i}}-D vector\footnote{See figure \ref{fig:1}.}) and \textit{E\textsubscript{j}} (\textit{l\textsubscript{j}}-D vector) are two input encoding vectors to the tensor fusion layer, the output \textbf{TFL}(\textit{E\textsubscript{i}}, \textit{E\textsubscript{j}}) is given by the following equation:
\begin{align}
    \textbf{TFL}(\textit{E\textsubscript{i}}, \textit{E\textsubscript{j}}) 
    = 
    \begin{pmatrix}
        \begin{bmatrix}
            \textit{E\textsubscript{i}} \\           
            1 
        \end{bmatrix} 
        \bigotimes
        \begin{bmatrix}
            \textit{E\textsubscript{j}} \\
            1
        \end{bmatrix}
    \end{pmatrix}
\end{align}
  \noindent where $\bigotimes$ is the operator for the outer product of two vectors. The constant 1 is appended to each of the input vectors \textit{E\textsubscript{i}} and \textit{E\textsubscript{j}} to include the individual features \textit{E\textsubscript{i}} and \textit{E\textsubscript{j}}.  \textbf{TFL}(\textit{E\textsubscript{i}}, \textit{E\textsubscript{j}}) is a (\textit{l\textsubscript{i}}+1)$\times$(\textit{l\textsubscript{j}}+1) dimensional matrix, which is flattened into a ((\textit{l\textsubscript{i}}+1)$\times$(\textit{l\textsubscript{j}}+1))-D vector. Since the tensor fusion layer is mathematically defined, it has no learnable parameters.
Within CombAtt, we define a set of pairs of input encodings \textit{TFI} = \{(\textit{TFI\textsubscript{11}}, \textit{TFI\textsubscript{12}}), (\textit{TFI\textsubscript{21}}, \textit{TFI\textsubscript{22}}), ..., (\textit{TFI\textsubscript{r1}}, \textit{TFI\textsubscript{r2}})\}, where \textit{TFIS} = \{\textit{TFI\textsubscript{11}}, \textit{TFI\textsubscript{12}}, \textit{TFI\textsubscript{21}}, \textit{TFI\textsubscript{22}}, ..., \textit{TFI\textsubscript{r1}}, \textit{TFI\textsubscript{r2}}\} is a subset of \textit{E}. We feed the tuples of \textit{TFI} to \textbf{TFL}, and obtain the output set of bi-modal encodings \textit{TFO} = \{\textit{TFO\textsubscript{1}}, \textit{TFO\textsubscript{2}}, ..., \textit{TFO\textsubscript{r}}\} where $ \textit{TFO\textsubscript{k}} = \textbf{TFL}(\textit{TFI\textsubscript{k1}}, \textit{TFI\textsubscript{k2}})$.
We define \textit{TFNS} = \textit{E} $\setminus$ \textit{TFIS}, the set difference of \textit{E} and \textit{TFIS}. The set of input encoding vectors to the attention fusion subnetwork is defined as \textit{MV} = \{\textit{MV\textsubscript{1}}, \textit{MV\textsubscript{2}}, ..., \textit{MV\textsubscript{s}}\} = \textit{TFO} $\cup$ \textit{TFNS}.


\subsubsection{\textbf{Attention fusion subnetwork}}
By using an attention mechanism, the network ``attends'' to the most relevant part of the input to generate the output. Networks with an attention mechanism usually perform better than their counterpart without attention. As not all modalities are equally relevant for the estimation of depression level, this motivates the introduction of an attention fusion subnetwork, as an extension of the work of \cite{poria2017multi}.

\begin{figure}
\includegraphics[scale = 0.9]{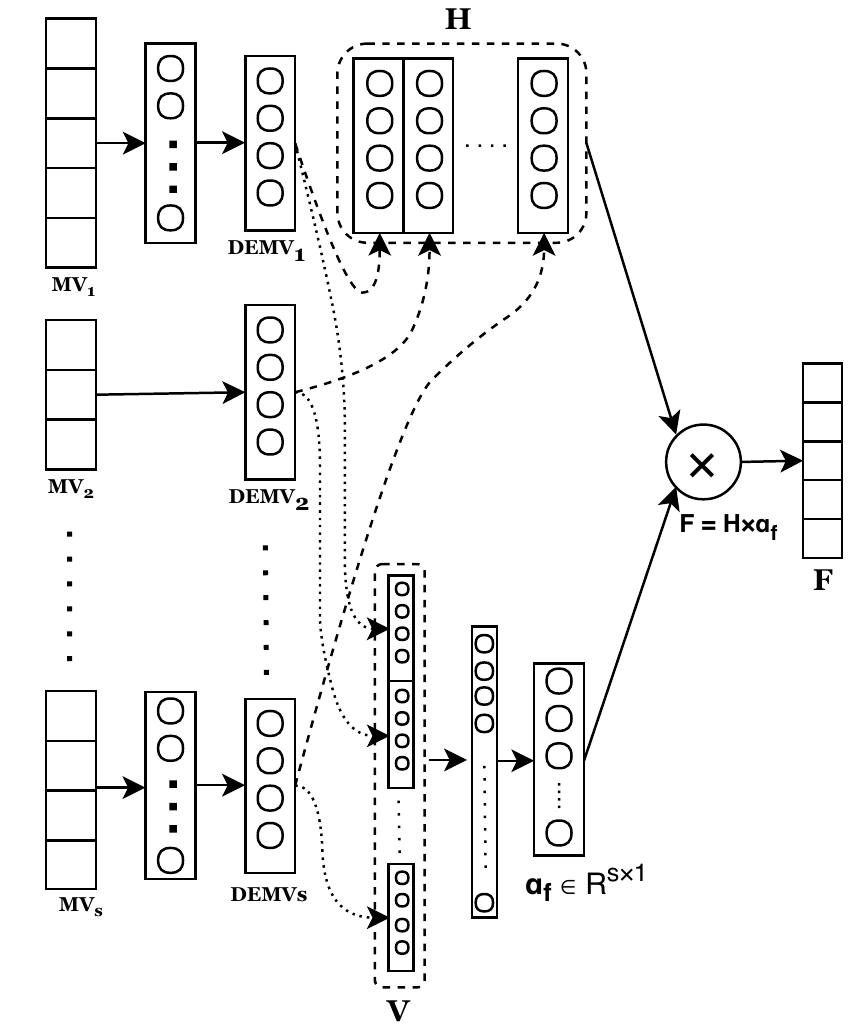}
\caption{Attention fusion subnetwork.}
\label{fig:2}
\end{figure}

The attention fusion subnetwork is shown in Figure \ref{fig:2}. The input to the attention fusion subnetwork is \textit{MV}, where the dimensionality of a vector \textit{MV\textsubscript{k}} $\in$ \textit{MV} is \textit{d\textsubscript{k}}. The first step of the architecture consists in giving the same dimension \textit{d} to all the elements of \textit{MV}. This is done using a stack of one or more dense layers. The resultant vectors are denoted by \textit{DEMV}, which is the set \{\textit{DEMV\textsubscript{1}}, \textit{DEMV\textsubscript{2}}, ..., \textit{DEMV\textsubscript{s}}\}. Then, all the elements of \textit{DEMV} are concatenated vertically into a vector \textit{V} and passed through a deep regression network, called the attention generation subnetwork and represented by \textbf{N\textsubscript{att}}. \textbf{N\textsubscript{att}} outputs a vector of attention values \textit{$\alpha$\textsubscript{f}} $\in$ $\mathbb{R}$\textsuperscript{s$\times$1}. Our attention generation subnetwork differs from \cite{poria2017multi} in the way that we let the deep regression network decide its parameters without any constraint for the generation of attention values. So, let \textit{W\textsubscript{att}} be the parameters of the attention generation subnetwork, \textbf{N\textsubscript{att}} is defined in equation \ref{eq:3}.

\begin{equation}
    \textit{$\alpha$\textsubscript{f}} = \textbf{N\textsubscript{att}} (\textit{V} : \textit{W\textsubscript{att}})
    \label{eq:3}
\end{equation}

\noindent In parallel, let \textit{H} = [\textit{DEMV\textsubscript{1}} $\mid$ \textit{DEMV\textsubscript{2}} $\mid$ ... \textit{DEMV\textsubscript{s}}] $\in$ $\mathbb{R}$\textsuperscript{d$\times$s} be the matrix obtained after the horizontal concatenation of the vectors in the set \textit{DEMV}. The fusion of the elements of \textit{DEMV} together with the attention values is performed as in equation \ref{eq:4}.

\begin{equation}
    \textit{F} = \textit{H.$\alpha$\textsubscript{f}}
    \label{eq:4}
\end{equation}

\noindent \textit{F} $\in$ $\mathbb{R}$\textsuperscript{d$\times$1} is the fusion vector that we feed to the PHQ-8 score regression subnetwork for the estimation of the depression level of a given patient.


\subsection{PHQ-8 Score regression subnetwork}
The PHQ-8 score regression subnetwork is a deep regression network conditioned on the fusion vector \textit{F}. The output \textit{F} of the attention fusion subnetwork is first concatenated to the gender of the patient\footnote{Note that gender plays an important role in the classification process of depression \cite{albert2015depression}.} and then fed to a few dense and dropout layers. The resultant vector is finally fed to a linear regression unit, which outputs the PHQ-8 score. So, let the PHQ-8 score regression subnetwork be denoted by \textbf{N\textsubscript{R}} and its parameters by \textit{W\textsubscript{R}}, the PHQ-8 score is estimated as in equation \ref{eq:5}.
\begin{equation}
    \textit{PHQ-8 score} = \textbf{N\textsubscript{R}} (\textit{F} : \textit{W\textsubscript{R}})
    \label{eq:5}
\end{equation}

In the following section, we present the gold standard DAIC-WOZ dataset that we use to perform our experiments.


\section{DAIC-WOZ depression dataset}
The DAIC-WOZ depression dataset\footnote{\url{http://dcapswoz.ict.usc.edu/}.} is part of a larger corpus, the Distress Analysis Interview Corpus \cite{gratch2014distress}, that contains clinical interviews designed to support the diagnosis of psychological distress conditions such as anxiety, depression, and post-traumatic stress disorder. These interviews were collected as part of a larger effort to create a computer agent that interviews people and identifies verbal and non-verbal indicators of mental illness. The data collected include audio and video recordings, and extensive questionnaire responses from the interviews conducted by an animated virtual interviewer called Ellie, controlled by a human interviewer in another room. The data has been transcribed and annotated for a variety of verbal and non-verbal features.

The dataset contains 189 sessions of interviews. We discarded a few interviews, as some of them were incomplete and others had interruptions. Each interview is recognized by a unique ID assigned to it. Each interview session contains a raw audio file of the interview session, a file containing the coordinates of 68 facial landmarks of the participant, a file containing HoG (Histogram of oriented Gradients) features of the face, two files containing head pose and eye gaze features of the participant, recorded over the entire duration of interview using a framework named OpenFace \cite{baltruvsaitis2016openface}, a file containing the continuous facial action units of the participant's face extracted using the facial action coding software CERT \cite{littlewort2011computer}, the COVAREP and formant feature files of the participant's voice extracted using a framework named COVAREP \cite{degottex2014covarep}, and a transcript file of the interview. All the features, leaving the transcript file, are time series data. A tabular description, showing some statistics of this dataset, is given in the appendix, in section 1.

Each row of the facial landmark file comprises the \textit{time stamp}, \textit{confidence}, \textit{detection success flag}, \textit{X}, \textit{Y} and \textit{Z} coordinates of each of the 68 facial landmarks. Each row of the head pose file comprises \textit{time stamp}, \textit{confidence}, \textit{detection success flag}, \textit{R\textsubscript{x}}, \textit{R\textsubscript{y}}, \textit{R\textsubscript{z}}, \textit{T\textsubscript{x}}, \textit{T\textsubscript{y}} and \textit{T\textsubscript{z}}. \textit{R\textsubscript{x}}, \textit{R\textsubscript{y}} and \textit{R\textsubscript{z}} are the head rotation coordinates (measured in radians), and \textit{T\textsubscript{x}}, \textit{T\textsubscript{y}} and \textit{T\textsubscript{z}} are the head position coordinates (measured in millimetres). The eye gaze feature file has rows that contain \textit{time stamp}, \textit{confidence}, \textit{detection success flag}, \textit{x\textsubscript{0}}, \textit{y\textsubscript{0}}, \textit{z\textsubscript{0}}, \textit{x\textsubscript{1}}, \textit{y\textsubscript{1}}, \textit{z\textsubscript{1}}, \textit{x\textsubscript{h0}}, \textit{y\textsubscript{h0}}, \textit{z\textsubscript{h0}}, \textit{x\textsubscript{h1}}, \textit{y\textsubscript{h1}} and \textit{z\textsubscript{h1}}. The gaze is represented by  4 vectors. The first two vectors (\textit{x\textsubscript{0}}, \textit{y\textsubscript{0}}, \textit{z\textsubscript{0}} and \textit{x\textsubscript{1}}, \textit{y\textsubscript{1}}, \textit{z\textsubscript{1}}) describe the gaze direction of both the eyes. The second two vectors (\textit{x\textsubscript{h0}}, \textit{y\textsubscript{h0}}, \textit{z\textsubscript{h0}} and \textit{x\textsubscript{h1}}, \textit{y\textsubscript{h1}}, \textit{z\textsubscript{h1}}) describe gaze in head coordinate space (if the eyes are rolled up, the vectors indicate 'up' even if the head is turned or tilted). Each row of the action units file comprises the \textit{time stamp}, \textit{confidence}, \textit{detection success flag}, and a few real numbers indicating the facial action unit. The data in these files were recorded at a frequency of 30Hz.

The COVAREP and formant feature files contain time series data. Each row of these files comprises 74 and 5 real numbers respectively, representing various features of the participant's voice or the virtual interviewer's voice. Both the features were recorded at 100Hz frequency. One of the features in both the files is a flag named \textit{VUV} (Voiced/Unvoiced), which denotes whether that segment is voiced or not. The DAIC-WOZ depression dataset manual advises  not to use those rows whose \textit{VUV} flag values are 0.

The transcript file contains the sentences spoken by Ellie and the participant. Each row of the file comprises \textit{start time}, the time at which the speaker starts speaking, \textit{stop time}, the time at which the speaker stops speaking, \textit{speaker}, denoting whether the speaker is Ellie or the participant, and \textit{value}, the exact sentence spoken by the speaker.

The training, development and test split files are provided with the dataset. The training and development split files comprise interview IDs, PHQ-8 binary labels, PHQ-8 scores, participant gender, and single responses to every question of the PHQ-8 questionnaire. The test split file comprises interview IDs and participant gender.


\section{Experiments and results}
This section is divided into three subsections. They are data preprocessing, experimental setup, and experimental results.


\subsection{Data preprocessing}
Different preprocessing techniques have been applied to different data modalities. We list them in the following subsections.

\subsubsection{Visual modalities preprocessing}
In the facial data modality, for every facial representation, we subtract the mean value of the Z-coordinate of the points from the Z-coordinates of all the points. This removes the bias along the Z-axis. Then, we normalize the points, so that the average distance to the origin is equal to 1. We calculate the distances between all possible pairs of points, and concatenate them with the normalized representation of the points. This results in a feature vector of size 2482 at each time step. In the head pose data modality, we rescale \textit{T\textsubscript{x}}, \textit{T\textsubscript{y}} and \textit{T\textsubscript{z}} by dividing them by 100. We downsample all the visual data modalities to 5Hz. We adopt a zero-tolerance strategy and discard all the time steps in all the visual modalities where the success flag is 0.  We do this to exclude the risk of introducing artefacts into the feature space. As the interviews of the participants are of different duration, we left-pad all the visual modality sequences with zero vectors along the temporal axis, to a common length of 10,000 time steps.

\subsubsection{Acoustic modalities preprocessing}
We discard those rows whose VUV flag values are 0. As this is an interview, the interviewer and the participant take turns to speak. We separate the participants COVAREP and formant features from the interview by using the start time and stop time given in the interview transcript file. We discard all those rows which belong to the turn of the participant where he/she has spoken for less than one second. We left-pad the COVAREP and formant features of all the participants with zeros along the time axis, to obtain a common length of 80,000 and 120,000 time steps respectively.

\subsubsection{Text modality preprocessing}
We collect only the participants utterances, and sort them according to their start times. As many of the participants have spoken colloquially, we formalize the utterances by replacing the contractions with the corresponding full words. Now, each utterance is encoded into a 512-D vector by using a pretrained Universal Sentence Encoder \cite{cer2018universal}. In this way, we construct time series data using the transcript file. We left-pad this data with zeros along the temporal axis, to obtain a common length of 400 time steps for all the participants.


\subsection{Experimental setup}
The hyperparameter details of each of the components of CombAtt network are described in the following sections. These hyperparameters were set empirically, after searching over a fairly vast hyperparameter space. The training procedure is provided in section 4 of the appendix. 

\subsubsection{Modality encoders}
In the Facial Landmark Encoder (\textbf{ME\textsubscript{FL}}), we use a LSTM layer with 256 memory cells. We feed the output of the LSTM layer to a dropout layer with a dropout rate of 0.3. The output of the dropout layer is passed through a 20 unit ReLU layer. The output of this hidden layer is again fed to a dropout layer of rate 0.3, before regressing the PHQ-8 score in the output layer. The facial landmark encoding, a 256-D vector obtained from \textbf{ME\textsubscript{FL}}, is denoted by \textit{E\textsubscript{FL}}.

The Head Pose Encoder (\textbf{ME\textsubscript{HP}}) contains a 2-layer LSTM, with 6 and 5 memory cells. The output of the second LSTM layer is fed to a dropout layer with a dropout rate of 0.166. The resultant vector is then concatenated to the gender of the participant, and passed through a 5 unit ReLU layer and a dropout layer with a rate of 0.25, before regressing the PHQ-8 score. The head pose encoding, a 5-D vector obtained from \textbf{ME\textsubscript{HP}}, is denoted by \textit{E\textsubscript{HP}}.


In the Eye Gaze Encoder (\textbf{ME\textsubscript{EG}}), we use a single LSTM layer with 64 memory cells. The output of the LSTM layer is concatenated to the gender of the participant, and passed through a couple of dense dropout layers, before regressing the PHQ-8 score. The dropout rates of the first and the second dropout layers are 0.2 and 0.125, respectively. The number of hidden units in the first and the second dense layers are 32 and 8, respectively. ReLU activation is used in the hidden dense layers. The eye gaze encoding, a 64-D vector obtained from \textbf{ME\textsubscript{EG}}, is denoted by \textit{E\textsubscript{EG}}.

The Action Units Encoder (\textbf{ME\textsubscript{AU}}) contains a single LSTM layer with 15 memory cells. The output of the LSTM layer is passed through a dropout layer of rate = 0.133. The resultant vector is appended to the gender of the participant, and is fed to a 6 unit ReLU layer. The output from this dense layer is fed to a dropout layer of rate 0.166, before regressing the PHQ-8 score in the output layer. The action units encoding, a 15-D vector obtained from \textbf{ME\textsubscript{AU}}, is denoted by \textit{E\textsubscript{AU}}.

The COVAREP and formant feature encoders (\textbf{ME\textsubscript{COV}} and \textbf{ME\textsubscript{FMT}}) contain single LSTM layers with 37 and 10 memory cells respectively. The outputs from the LSTM layers of both the networks are passed through a dropout layer whose rate is 0.2 and 0.25 in COVAREP and formant feature encoders, respectively. Gender of the participant is appended to the output of this dropout layer. The resultant vectors pass through a stack of dense-dropout layers, and the output of this stack is used for regressing the PHQ-8 score in the output layer. The rate of the second dropout layer is 0.2 and 0.25 in COVAREP and formant feature encoders, respectively. The dense layer in \textbf{ME\textsubscript{COV}} has 15 hidden units, and the dense layer in \textbf{ME\textsubscript{FMT}} has 6 units. ReLU activation is used in the hidden layers of both the encoders. The COVAREP and formant feature encodings (37-D and 10-D vectors) obtained from these networks are denoted by \textit{E\textsubscript{COV}} and \textit{E\textsubscript{FMT}} respectively.

The transcript encoder (\textbf{ME\textsubscript{TR}}) contains a single LSTM layer with 200 memory cells. The sum of the outputs from all the LSTM time steps is passed to a dropout layer of rate = 0.3. The resultant vector is fed to a 60 unit ReLU layer. The output from this dense layer is fed to a dropout layer of rate 0.3, before regressing the PHQ-8 score in the output layer. The transcript encoding, a 200-D vector obtained from \textbf{ME\textsubscript{TR}}, is denoted by \textit{E\textsubscript{TR}}.

\subsubsection{\textbf{Tensor fusion layer}}
We pair the visual and acoustic modalities into three groups - (\textit{E\textsubscript{FL}}, \textit{E\textsubscript{HP}}), (\textit{E\textsubscript{AU}}, \textit{E\textsubscript{EG}}), and (\textit{E\textsubscript{COV}}, \textit{E\textsubscript{FMT}}), and pass them to the tensor fusion layer to obtain the outputs \textit{TFO\textsubscript{FL$\times$HP}} (1542-D vector), \textit{TFO\textsubscript{AU$\times$EG}} (1040-D vector), and \textit{TFO\textsubscript{COV$\times$FMT}} (418-D vector) respectively. These three outputs, along with \textit{E\textsubscript{TR}}, are passed on to the attention fusion subnetwork.

\subsubsection{\textbf{Attention fusion subnetwork}}\label{subnet}
We first shorten \textit{TFO\textsubscript{FL$\times$HP}},  \textit{TFO\textsubscript{AU$\times$EG}}, \textit{TFO\textsubscript{COV$\times$FMT}}, and elongate \textit{E\textsubscript{TR}} to a length of 450. \textit{TFO\textsubscript{FL$\times$HP}} is shortened by feeding it to a stack of 600 unit ReLU and 450 unit ReLU layers, to obtain \textit{DEMV\textsubscript{FL$\times$HP}}. \textit{TFO\textsubscript{AU$\times$EG}} is shortened by feeding it to a stack of 570 unit ReLU and 450 unit ReLU layers, to obtain \textit{DEMV\textsubscript{AU$\times$EG}}. \textit{TFO\textsubscript{COV$\times$FMT}} is shortened by feeding it to a 450 unit ReLU layer, to obtain \textit{DEMV\textsubscript{COV$\times$FMT}}. \textit{E\textsubscript{TR}} is elongated by feeding it to a stack of 315 unit ReLU and 450 unit ReLU layers, to obtain \textit{DEMV\textsubscript{TR}}. The attention generation subnetwork takes the vertical concatenation of a subset \textit{SM}\footnote{SM is the set of modalities which are being fused.} $\in$ \{\textit{DEMV\textsubscript{FL$\times$HP}}, \textit{DEMV\textsubscript{AU$\times$EG}}, \textit{DEMV\textsubscript{COV$\times$FMT}}, \textit{DEMV\textsubscript{TR}}\} as input, and feeds it to a 300 unit tanh layer. The output of this hidden layer is passed on to a softmax layer which outputs the attention values. The number of hidden units in this softmax layer is determined by the length of the subset modalities \textit{SM}. 

To see how the different combinations of acoustic, text and visual modalities help in the estimation of PHQ-8 score, and to see the effect of using attention for fusion of modalities, we trained and tested the following networks. In the network \textbf{CombAtt\textsubscript{a$\times$v}}, we fuse \textit{DEMV\textsubscript{COV$\times$FMT}}, \textit{DEMV\textsubscript{FL$\times$HP}} and \textit{DEMV\textsubscript{AU$\times$EG}} (the audio and visual modality encodings), and pass them on to the regression subnetwork. In the network \textbf{CombAtt\textsubscript{t$\times$v}}, we fuse \textit{DEMV\textsubscript{TR}}, \textit{DEMV\textsubscript{FL$\times$HP}} and \textit{DEMV\textsubscript{AU$\times$EG}} (the text and visual modality encodings), and feed them to the regression subnetwork. In the network \textbf{CombAtt\textsubscript{a$\times$t}}, we fuse \textit{DEMV\textsubscript{COV$\times$FMT}} and \textit{DEMV\textsubscript{TR}} (acoustic and text modality encodings), and feed them to the regression subnetwork. In the network \textbf{CombAtt\textsubscript{a$\times$t$\times$v}}, we fuse \textit{DEMV\textsubscript{COV$\times$FMT}}, \textit{DEMV\textsubscript{TR}}, \textit{DEMV\textsubscript{FL$\times$HP}} and \textit{DEMV\textsubscript{AU$\times$EG}} (the audio, text and visual modality encodings), and pass them on to the regression subnetwork. In the network \textbf{CombAtt\textsubscript{attentionless}}, we concatenate \textit{DEMV\textsubscript{COV$\times$FMT}}, \textit{DEMV\textsubscript{TR}}, \textit{DEMV\textsubscript{FL$\times$HP}} and \textit{DEMV\textsubscript{AU$\times$EG}} vertically, and pass them on to a stack of dense-dropout layers. The dense layer has 300 hidden units, and uses ReLU as its activation. The dropout rate of the dropout layer is 0.25. The output from this stack is appended to the gender, and is passed on to a regression unit, which regresses the PHQ-8 score.

\subsubsection{\textbf{Regression subnetwork}}
The input of this network is the concatenation of the fused modalities with the gender of the participant, and is fed to a couple of dense dropout layers. The first and second dense layers have 310 and 83 hidden units, respectively, and use ReLU as the activation function. The corresponding dropout layers have rates of 0.25 and 0.2, respectively. The output of the second hidden layer is fed to a regression unit, which regresses the PHQ-8 score (a integer value between 0 and 24).


\subsection{Experimental results}
We first compare the results of the networks, which use different combinations of modalities, as defined in sections 5.2.1 and 5.2.3, in Table \ref{tab:1}. In particular, we use three evaluation metrics: root mean squared error (RMSE), mean absolute error (MAE) and explained variance score (EVS).
 
\begin{table}[h!]
  \begin{center}
    \label{tab:table1}
    \begin{tabular}{|l|l|l|l|}

    \hline
    \textbf{Network} & \textbf{RMSE} & \textbf{MAE} & \textbf{EVS}\\
    \hline
    \textbf{ME\textsubscript{FL}} & 6.24 & 5.30 & 0.12\\
    \textbf{ME\textsubscript{HP}} & 6.45 & 5.24 & 0.08\\
    \textbf{ME\textsubscript{EG}} & 6.57 & 5.45 & 0.04\\
    \textbf{ME\textsubscript{AU}} & 6.53 & 5.06 & 0.18\\
    \hline
    \textbf{ME\textsubscript{COV}} & 6.60 & 5.71 & 0.03\\
    \textbf{ME\textsubscript{FMT}} & 6.65 & 5.66 & 0.01\\
    \hline
    \textbf{ME\textsubscript{TR}} & \textbf{4.80} & \textbf{3.74} & \textbf{0.48}\\
    \hline
    \hline
    \textbf{CombAtt\textsubscript{a$\times$v}} & 5.25 & 3.89 & 0.39\\
    \textbf{CombAtt\textsubscript{t$\times$v}} & 5.11 & 3.65 & 0.48\\
    \textbf{CombAtt\textsubscript{a$\times$t}} & 4.64 & 3.65 & 0.57\\
    \hline
    \textbf{CombAtt\textsubscript{attentionless}} & 4.69 & 3.66 & 0.50\\
   \textbf{CombAtt\textsubscript{a$\times$t$\times$v}} & \textbf{4.14} & \textbf{3.07} & \textbf{0.62}\\
    \hline
    \hline
    \textbf{AW\textsubscript{behavioural}} & 5.54 & 4.73 & NA\\
    \textbf{MMD} & 4.65 & 3.98 & NA \\
    \textbf{VFSC\textsubscript{semantic}} & \textbf{4.46} & \textbf{3.34} & NA \\
    \hline
    \end{tabular}
        \caption{Results for combinations and state-of-the-art.}
        \label{tab:1}
  \end{center}
\end{table}

It is clear that the fusion of all modalities helps in better estimation of depression level, compared to any other combination (\textbf{CombAtt\textsubscript{a$\times$t$\times$v}} outperforms all the other networks). Among the modality encoders, \textbf{ME\textsubscript{TR}} gives the most accurate estimate, indicating that verbal input plays a predominant role in estimating depression level, which is confirmed by therapists' experience. Results also show the benefits of the attention mechanism, which allows to take into account the weight of each modality in the classification process. Note that the results obtained by \textbf{CombAtt\textsubscript{a$\times$t$\times$v}} are statistically significant when compared to all other configurations.


We also compare the performance of the CombAtt network with three approaches, that can be considered as the state-of-the-art within this task:

\noindent $\textbf{AW\textsubscript{behavioural} : }$ This is the winning approach \cite{stepanov2017depression} in AVEC 2017 \cite{ringeval2017avec} depression sub-challenge. The authors use feature extraction methods on acoustic and text features, and recurrent neural network on visual features. It is the current state-of-the-art on the test split of DAIC-WOZ. The authors develop four models, but we compare our approach with the best of these four models (\textbf{AW\textsubscript{behavioural}}). 


\noindent $\textbf{MMD : }$ In this approach \cite{yang2017multimodal}, the authors propose a multimodal fusion framework composed of deep convolutional neural network (DCNN) and deep neural network (DNN) models. The framework considers acoustic, text and visual streams of data. For each modality, handcrafted feature descriptors are fed to a DCNN that learns high-level global features with compact dynamic information. Then, the learned features are fed to a DNN to predict the PHQ-8 scores. For multimodal fusion, the estimated PHQ-8 scores from the three modalities are integrated in another DNN to obtain the final PHQ-8 score.

\noindent $\textbf{VFSC\textsubscript{semantic} : }$ In this approach \cite{williamson2016detecting}, the authors derive the biomarkers from visual, acoustic and text modalities. The authors define semantic context indicators, which use the provided transcripts to infer a subject’s status with respect to four conceptual classes. The semantic context feature is the sum of points accrued from all four indicators. This approach is the state-of-the-art on the official development split of DAIC-WOZ.

The results reported in Table \ref{tab:1} show that CombAtt outperforms all state-of-the-art approaches. In particular, CombAtt achieves an increase of 7.17\% on RMSE and 8.08\% on MAE over \textbf{VFSC\textsubscript{semantic}} (the best performing state-of-the-art). On performing statistical significance tests, we found that the improvements are statistically significant.

\section{Conclusion}
In this paper, we introduce an attention-based fusion network (CombAtt) for the estimation of PHQ-8 score. Experimental results show that (1) fusing all modalities helps in giving the closest estimation of depression level, (2) the text modality plays an important role in the regression process, and (3) CombAtt outperforms previous state-of-the-art approaches. However, there is still a great margin for improvement. Indeed, as stated in \cite{valstar2016avec}, a baseline classifier that constantly predicts the mean score of depression provides an RMSE=5.73 and an MAE=4.74. In our future work, we propose to (1) study recent multi-task learning architectures such as \cite{sanh2018hierarchical} and (2) dig deeper into high-level text representations such as \cite{devlin2018bert}.

\bibliographystyle{acl_natbib.bst}
\bibliography{naaclhlt2019.bbl}

\end{document}